# Enhancing Keyphrase Extraction from Academic Articles with their Reference Information

Chengzhi Zhang*, Lei Zhao, Mengyuan Zhao, Yingyi Zhang

Department of Information Management, Nanjing University of Science and Technology, Nanjing 210094, China

**Abstract**: With the development of Internet technology, the phenomenon of information overload is becoming more and more obvious. It takes a lot of time for users to obtain the information they need. However, keyphrases that summarize document information highly are helpful for users to quickly obtain and understand documents. For academic resources, most existing studies extract keyphrases through the title and abstract of papers. We find that title information in references also contains author-assigned keyphrases. Therefore, this article uses reference information and applies two typical methods of unsupervised extraction methods (TF*IDF and TextRank), two representative traditional supervised learning algorithms (Naïve Bayes and Conditional Random Field) and a supervised deep learning model (BiLSTM-CRF), to analyze the specific performance of reference information on keyphrase extraction. It is expected to improve the quality of keyphrase recognition from the perspective of expanding the source text. The experimental results show that reference information can increase precision, recall, and $F_1$ of automatic keyphrase extraction to a certain extent. This indicates the usefulness of reference information on keyphrase extraction of academic papers and provides a new idea for the following research on automatic keyphrase extraction.

**Keywords**: Keyphrase extraction, References, Unsupervised learning, Supervised learning

## 1 Introduction

The explosive growth of literature has far exceeded users' reading and understanding ability, which makes it extremely difficult for users to select the information they need. The Keyphrase, usually consisting of one or more salient words, represents the significant content of a document, which helps users quickly understand the topic of a document [1]. Moreover, keyphrases are widely used in tasks of research topic analysis [2], knowledge mapping [3], and domain analysis [4]. Selecting high-quality keyphrases from a document is critical for improving the quality of these tasks. Although manual keyphrase extraction is highly accurate, it is time and labor-consuming. Nowadays, automatic keyphrase extraction has become an important task in natural language processing, due to it can improve the efficiency of keyphrase extraction. Automatic keyphrase extraction is designed to extract important words or phrases that reflect the topics of a document [5]. Nowadays, automatic keyphrase extraction has been widely applied in many fields such as information retrieval [6, 7] and text classification [8, 9].

In academic resources, keyphrases offer the core information of academic papers and are convenient for researchers to search papers. However, not all the authors provide keyphrases [10]. For example, there is no keyphrase provided by articles published in *PLOS ONE*. Thus, some studies selected keyphrases based on the title and abstract which mostly represent the main content of papers. Nevertheless, extracting keyphrases from titles and abstracts may have a data sparsity problem. Given a large amount of full-text information, machine learning based on full-text requires lots of computing space, which has high processing cost and low extraction efficiency. In addition, considering that the reference is also an influential part of academic papers, and provides research background and theoretical basis for the paper, there is a close correlation between research content and the reference [11]. To calculate the coverage of keyphrases in the title, abstract, and reference titles, we carried out a preliminary investigation on the SemEval-2010 Task 5 dataset (SemEval-2010)[1], the PubMed dataset provided by Schutz [12], and our self-built dataset (LIS-2000). Table 1 shows the coverage of keyphrases on the training and test sets of SemEval-2010, the dataset of PubMed and the self-built dataset of LIS-2000. It can be seen that the reference title not only includes keyphrases in titles and abstracts, but also contains keyphrases that do not appear in titles and abstracts. Accordingly, we try to use reference information to expand the data context and explore whether the reference information can help identify more correct keyphrases.

* Corresponding Author, E-mail: zhangcz@njust.edu.cn.

Table 1: Coverage of keyphrases assigned by their authors in the titles, abstracts and reference titles on the dataset of SemEval-2010, PubMed and LIS-2000

| Dataset / Coverage | SemEval-2010 | PubMed | LIS-2000 |
|---|---|---|---|
| **Titles and abstracts** | 52.85% | 63.65% | 56.36% |
| **Reference titles** | 47.25% | 59.74% | 51.90% |
| **Reference titles (not appear in titles and abstracts)** | 13.64% | 14.02% | 12.34% |

Existing researches on automatic keyphrase extraction from academic papers include unsupervised and supervised extraction methods. Unsupervised extraction methods extract keyphrases directly from the text without annotation corpus, such as TF*IDF [13] and TextRank [14]. Keyphrase extraction methods based on supervised methods need to train models with manually annotated corpus, and then use the obtained model to predict keyphrases from another dataset, such as Naïve Baye (NB) [15], Support Vector Machine (SVM) [16], Conditional Random Field (CRF) [17] and BiLSTM-CRF [18]. Based on existing researches, we utilize both unsupervised and supervised methods to extract keyphrases from academic papers. For unsupervised methods, we select TF*IDF and TextRank model. For supervised methods, we select two traditional supervised models, i.e., NB and CRF model and a supervised deep learning model (BiLSTM-CRF).

The contributions of our article are two folds: first, different from most methods extract keyphrases from titles and abstracts, this article tries to utilize reference titles to enrich the context and thereby explore the influence of reference information on keyphrase extraction. Second, we use five models, i.e., TF*IDF, TextRank, NB, CRF, and BiLSTM-CRF, and compare keyphrase extraction performance with or without reference titles on each method. Moreover, we conduct qualitative experiments and analyze the specific influence of reference titles on keyphrase extraction.

In the rest of this article, Section 2 is the relevant work of keyphrase extraction methods. We describe data collection, research methods, and evaluation metrics in Section 3. Then, the experimental results of different extraction methods are analyzed in Section 4. Section 5 and 6 are the discussion and summarization.

[1] https://semeval2.fbk.eu/semeval2.php?location=data

All the data and source code of this paper are freely available at github website: https://github.com/chengzhizhang/Keyphrase_Extraction.

## 2 Related Work

Automatic keyphrase extraction methods can be divided into two categories, i.e., unsupervised method and supervised method. Supervised methods require a training corpus that is annotated manually while unsupervised methods do not need [19].

### 2.1 Automatic Keyphrase Extraction Methods Based on Unsupervised Learning

Unsupervised keyphrase extraction methods can directly extract keyphrase by ranking candidate keyphrase [20]. TF*IDF proposed by Salton [13] is the common baseline in the statistics-based methods, which calculate word weights by word frequency and then rank words by their weights. But TF*IDF ignores features of words, such as the occurrence of important words and word location, so many scholars constantly put forward improved methods. Basili et al. proposed Inverse Word Frequencies (IWF) which represents the logarithm of the ratio between the total number of all words and the number of word t in the corpus, taking into account the distribution of words in the corpus [21]. Liu et al. incorporated part-of-speech (POS) information, word clustering, and sentence salience score as additional knowledge [22]. Their experimental results showed that these improved TF * IDF methods had a better performance [19, 20]. In addition, some approaches extract keyphrases linked to the topics of a document, i.e., Latent Semantic Analysis (LSA) [23], Probabilistic Latent Semantic Indexing (pLSI) [24], Latent Dirichlet Allocation (LDA) [25], and Topic Distilling with Compressive Sensing (TDCS) [26]. Wang et al. proposed a topic-aware sequence-to-sequence (seq2seq) model recently, which outperformed state-of-the-art methods due to their model allowed the joint learning of latent topic representations [27]. Beyond that, TextRank proposed by Mihalcea and Tarau was the first graph-based keyphrase extraction method [14]. Based on traditional TextRank, Bellaachia and Al-Dhelaan introduced new features to improve keyphrase extraction on Twitter including edge weight and word weight [28]. Moreover, Yang et al. proposed a network model considering the influence of sentences and functions based on WordNet [29].

From the existing research, we can see that scholars have been considering adding more impact factors to optimize unsupervised keyphrase extraction methods. Empirical studies conducted by Hasan and Ng showed that TF*IDF has strong performance and can be used as a strong baseline standard in keyphrase extraction methods [30]. TextRank is proposed based on the Google PageRank algorithm [31] and has achieved good results, which takes into account the co-occurrence relationship between words. Thus, this article chooses these two unsupervised methods to extract keyphrases.

### 2.2 Automatic Keyphrase Extraction Methods Based on Supervised Learning

Automatic keyphrase extraction methods based on supervised learning train a model from a labeled training set, and then use this model to predict labels for a new dataset [20]. The task of keyphrase extraction can be transformed into a classification problem. For example, Witten et al. put forward a new Keyphrase Extraction Algorithm (KEA), which considered the TF*IDF score and the first occurrence of each candidate phrase and used Naïve Bayes for keyphrase extraction [32]. Turney proposed the GenEx algorithm which consisted of the Genitor genetic algorithm and the Extractor keyphrase extraction algorithm [33]. Zhang et al. also transformed the keyphrase extraction problem into a classification problem and adopted SVM as the classifier, taking into account global context features and local context features [16]. In addition, keyphrase extraction also is considered an annotation problem, such as CRF models. Zhang et al. tried to use the title, abstract, heading of paragraph or section of academic papers to improve keyphrase extraction, which verified the effectiveness of additional information and outperformed SVM, multiple linear regression model in keyphrase extraction [34]. In addition, Zhang et al. exploited a neural network keyphrase extraction model combined with comment content based on Twitter datasets, which enriched target posts through encoding comment content [35]. Zhang et al. utilized human reading time extracted from an open-source eye-tracking dataset to enhance keyphrase extraction with neural network models [34].

Based on previous research, the supervised machine learning model has achieved superior performance in keyphrase extraction tasks. For classification tasks, the NB algorithm is adopted in the article to extract keyphrases, and the Naïve Bayes classifier is used for model training considering the TF*IDF score and the first occurrence location. Regarding sequence labeling tasks, some scholars compared the CRF model with Hidden Markov Model (HMM), which found that CRF is more suitable for processing sequence labeling tasks than HMM [37]. Moreover, the CRF model can arbitrarily select features and normalize them globally. In the deep learning model, BiLSTM-CRF is widely used in keyphrase extraction tasks and has been proven to have superior performance [38]. Therefore, this paper also chooses this model to enhance the reliability of the conclusions.

In the relevant research of automatic keyphrase extraction for academic papers, most scholars take the title and abstract as the corpus to conduct keyphrase extraction. In addition, Ma & Hou attempted to use a predefined lexicon to extract keyphrases from titles of the article and the corresponding references, and compared the results of manual annotation and automatic annotation, to verify the feasibility of extracting keyphrases from reference title information [11]. Some scholars also tried to obtain keyphrases from the perspectives of reference keyphrases [39], citation content [40], or the background knowledge from Wikipedia [41]. In recent years, many abstractive methods have been proposed, which have produced meaningful keyphrases [42]. But this article aims at introducing reference title information to enrich the context, which uses unsupervised models, i.e., TF*IDF and TextRank, and supervised models, i.e., NB, CRF, and BiLSTM-CRF model, as keyphrases extraction methods. The article not only verifies that reference information could improve the effect of keyphrase extraction, but also elaborates on the concrete manifestation of the influence of reference information on keyphrase extraction.

## 3 Methodology

In order to explore the influence of reference information on keyphrase extraction in academic papers, we first analyze the logical structure of the paper and extract the title, abstract, keyphrases assigned by authors, introduction, conclusion, first sentence of paragraph, last sentence of paragraph, reference title and full text. Then, we take the title and abstract of the paper as the basic corpus and add different logical structure texts to construct multiple corpora, especially focusing on the use of reference information. Next, data preprocessing is carried out, including lemmatization, symbol removal, stop-words filtering and POS tagging. Finally, we utilize TF*IDF, TextRank, NB algorithm, CRF and BiLSTM-CRF model to extract keyphrases from the constructed multiple corpora automatically and finally evaluate and compare the extraction results. Figure 1 shows the research framework for automatic keyphrase extraction with their reference information.

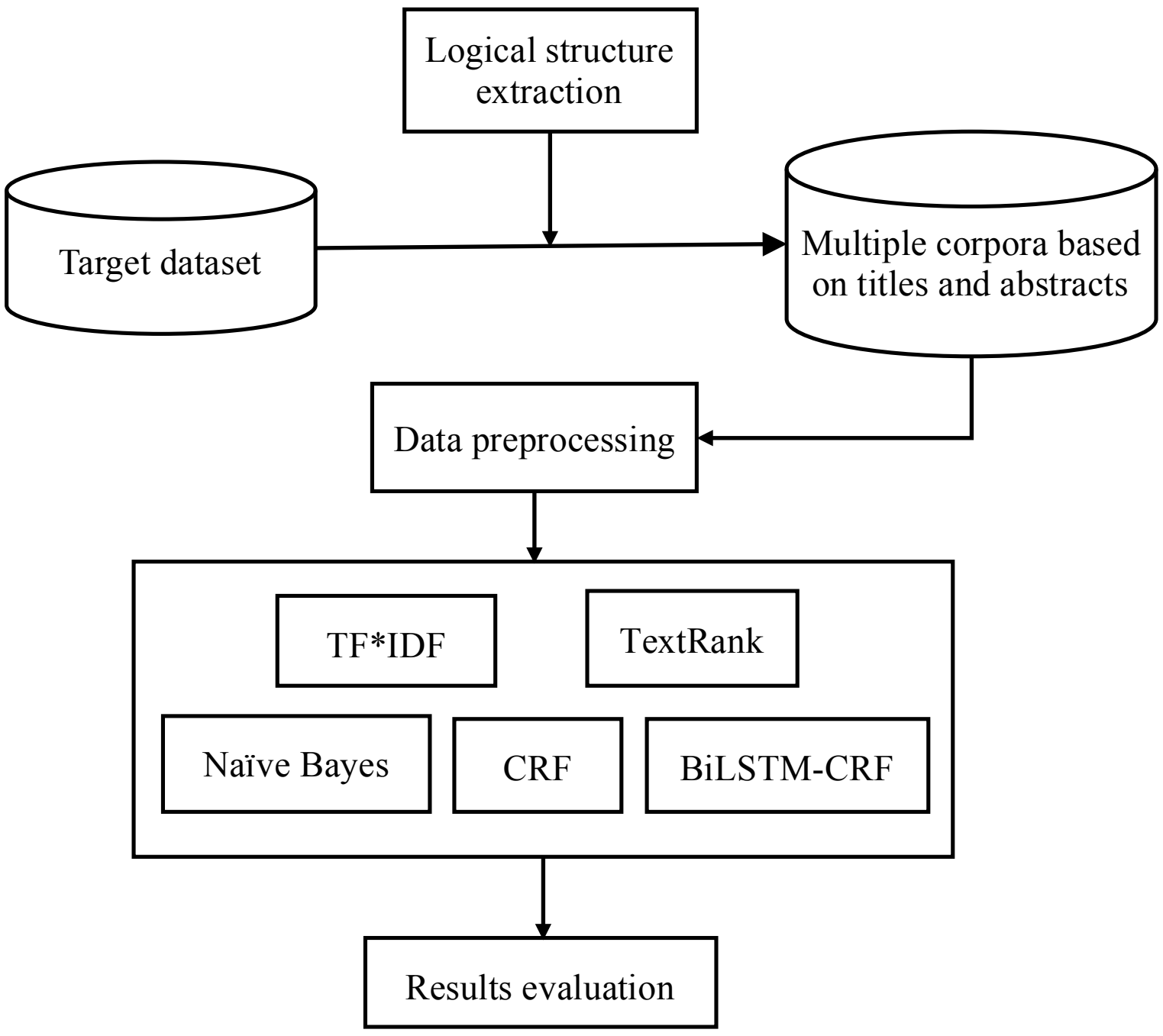

Figure 1: Framework for automatic keyphrase extraction with their reference information

### 3.1 Corpus Construction for Keyphrase Extraction

Considering the feasibility of obtaining reference information and full text, three datasets are selected in this paper. The first is the SemEval-2010 Task 5 dataset (SemEval-2010), the second is the PubMed dataset, and the last is the self-built dataset named LIS-2000. The following is a brief description of these datasets.

**(1) SemEval-2010 dataset**

This dataset contains 144 full-text academic papers with keyphrases and references as the training set, as well as 100 full-text academic papers with keyphrases and references as the test set. Since the small difference in the number of papers contained in the training and test sets, we combined and re-divided the 244 papers in the experiment.

**(2) PubMed dataset**

This dataset is from Schutz [12] containing 1323 academic papers with titles, abstracts, keyphrases, reference information and full text, which are stored in XML files. By parsing the XML file, we found that there were 7 papers without full-text content, so we removed them. Finally, the dataset contains 1316 academic papers.

**(3) LIS-2000 dataset**

This dataset is our self-built dataset, which contains 2000 papers with titles, abstracts, keyphrases, reference information and and full text. These papers are randomly selected from scientific literatures of five library and information science journals (LIS), i.e., Aslib Journal of Information Management, Journal of Documentation, Library Hi Tech, Online Information Review, and The Electronic Library.

Based on the above datasets, we use the methods of rule matching and manual annotation to analyze the logical structure of the paper, and extract the title, abstract, introduction, conclusion, first sentence of paragraph, last sentence of paragraph, reference title, and full text. Then, we divide each dataset into training and test sets according to the ratio of 8:2. Table 2 shows the number of samples and coverage of keyphrases assigned by the authors in the training and test sets of SemEval-2010, PubMed and LIS-2000. It can be seen that the coverage of keyphrases in different logical structures on the training and test sets is similar to the overall coverage of keyphrases, which indicates that the division of dataset is unbiased. For unsupervised methods, we conduct experiments using the test set. In supervised methods, we use the training set to obtain the model, and then use the test set to evaluate its performance.

Table 2: The number of samples and coverage of keyphrases assigned by the authors in the training and test sets of SemEval-2010, PubMed and LIS-2000

| Dataset / Item | SemEval-2010 | | | PubMed | | | LIS-2000 | | |
|---|---|---|---|---|---|---|---|---|---|
| | **All** | **Train** | **Test** | **All** | **Train** | **Test** | **All** | **Train** | **Test** |
| **Number of samples** | 244 | 196 | 48 | 1316 | 1052 | 264 | 2000 | 1600 | 400 |
| title | 27.48% | 26.85% | 30.29% | 32.38% | 33.06% | 29.67% | 28.97% | 29.12% | 28.39% |
| abstract | 47.78% | 46.82% | 52.00% | 59.98% | 60.46% | 58.08% | 54.37% | 54.44% | 54.13% |
| introdcution | 64.38% | 63.16% | 69.71% | 61.26% | 61.90% | 58.71% | 55.19% | 55.17% | 54.32% |
| conclusion | 48.41% | 48.25% | 49.71% | 44.66% | 44.55% | 45.14% | 44.18% | 44.50% | 42.34% |
| First sentence of paragraph | 64.59% | 63.42% | 69.71% | 65.84% | 66.47% | 63.33% | 60.83% | 61.13% | 58.74% |
| Last sentence of paragraph | 61.52% | 60.70% | 64.57% | 58.46% | 58.94% | 56.54% | 55.49% | 55.52% | 54.37% |
| reference title | 47.25% | 46.82% | 49.14% | 59.74% | 60.46% | 56.89% | 51.90% | 52.17% | 50.83% |
| full text | 78.33% | 77.82% | 80.57% | 81.67% | 82.26% | 79.36% | 75.01% | 75.14% | 73.67% |
| **Total number of keyphrases** | **946** | **771** | **175** | **7104** | **5675** | **1429** | **10104** | **8068** | **2036** |

To understand the importance of each section in papers, we analyze the coverage of keyphrases assigned by authors in different logical structures. The results in Table 2 show that the reference title contains more keyphrases assigned by authors, which is equivalent to the number of keyphrases in the abstract. Besides, according to the results in Table 1, the 13.64%, 14.02% and 12.34% of keyphrases is only appearing in reference titles and not appearing in titles and abstracts in SemEval-2010, PubMed and LIS-2000, respectively. It indicates that adding reference title information could provide more correct keyphrases. Therefore, reference information is helpful for the automatic keyphrase identification in theory. In order to explore the influence of reference information on the keyphrase extraction task, we first take the title and abstract of the paper as the basic corpus. Then, texts with different logical structures are added to the basic corpus to construct multiple corpora. Finally, the keyphrase extraction model is used to evaluate the performance of extraction using the constructed multiple corpora, especially to compare the influence of reference information added.

### 3.2 Data Preprocessing

The data preprocessing has five steps including tokenizing, part-of-speech tagging, stop-word filtering, symbol removal, and lemmatization, which is mainly implemented by NLTK[2].

i) Tokenizing: Tokenizing is to separate words and punctuation marks with spaces.

ii) Part-of-speech tagging: it is the process of labeling each word with a corresponding part-of-speech including noun, adjective, adverb, verb, etc.

iii) Stop-words filtering: in English text, stop words are generally adverbs, conjunctions, and prepositions, such as "a, the, in, and, is ...", which have no concreted meaning and interfere with keyphrase recognition, so that these stop words are filtered based on dictionary.

[2] http://www.nltk.org/

iv) Symbol removal: the punctuation marks or special symbols are deleted, such as ", ", "#", ":", etc.

v) Lemmatization: lemmatization is to deform a word to a general form. Considering the words in the English corpus have different forms, lemmatization is necessary to unify the form of the word.

### 3.3 Keyphrase Extraction Methods

The article chooses two kinds of extraction methods to explore the contribution of reference information to automatic keyphrase extraction. One is based on unsupervised learning containing TF*IDF and TextRank, the other is based on supervised learning including NB, CRF, and BiLSTM-CRF.

#### 3.3.1 Automatic Keyphrase Extraction Methods Based on Unsupervised Learning

**(1) TF*IDF**

TF*IDF is a statistical method proposed by Salton in 1988 to evaluate the importance of a word to the document[13], which considers the frequency of a word appearing in the document. The calculation formula is:

$$\mathrm{TF} * \mathrm{IDF} = \frac{Freq(Word)}{\max_Freq} \times log_2 \frac{N+1}{n+1} \quad (1)$$

where $Freq(Word)$ represents the number of times a word appears in the document, $\max_\mathrm{Freq}$ denotes the maximum number of times a word appears in the document, $N$ is the total number of documents in the whole document set, $n$ indicates the number of documents containing this word. For a word, the larger its TF*IDF value, the more times the word appears in the document, and the fewer documents in the whole dataset that contain the word, which indicates that the word is more important to the document.

**(2) TextRank**

TextRank is an unsupervised keyphrase extraction method based on word graph models, which constructs the graph model based on the semantic relationship between adjacent words [14]. Firstly, according to the POS tags, nouns, verbs, and adjectives are selected as candidate keyphrases to form candidate keyphrase set $K = \{W_1, W_2, \cdots, W_m\}$. And then candidate keyphrase graph $\mathrm{G} = (\mathrm{V}, \mathrm{E})$ is constructed, where $\mathrm{V}$ represents the set of nodes in candidate keyphrase set $K$. $\mathrm{E}$ denotes the set of edges, which is added between the nodes that co-occur within a window of L words. And the weight value of each node is calculated through the recursive formula of TextRank until convergence. The formula is:

$$\mathrm{Score}(V_i) = (1-d) + d \times \sum_{V_i \in In(V_i)} \frac{w_{ji}}{\sum_{V_k \in Out(V_j)} w_{jk}} Score(V_j) \quad (2)$$

where $V_i, V_j$ belong to candidate keyphrase set $K$; $w_{ji}$ is the weight value of edges; $In(V_i)$ is the node set pointing to the node $V_i$; $Out(V_i)$ is the node set pointed out by the node $V_i$; $d$ is the damping coefficient, representing the probability of pointing from one node to other nodes in the graph with a value range from 0 to 1. We set it as 0.85 in our experiment [14].

**(3) Candidate Keyphrases Combination**

The weight value of each word in the document can be calculated through TF*IDF and TextRank algorithms. But considering the number of keyphrases assigned by authors maybe 1, 2, 3, or more in the academic papers, we count the number of keyphrases annotated by the author in the corpus, which is presented in Table 3.

Table 3: Word number distribution of keyphrases assigned by the author

| Word Number \ Dataset | Number of keyphrases | | | Proportion | | |
|---|---|---|---|---|---|---|
| | **SemEval-2010** | **PubMed** | **LIS-2000** | **SemEval-2010** | **PubMed** | **LIS-2000** |
| 1 | 273 | 3070 | 3178 | 28.86% | 43.24% | 31.45% |
| 2 | 504 | 2689 | 5634 | 53.28% | 37.87% | 55.76% |
| 3 | 130 | 905 | 923 | 13.74% | 12.75% | 9.13% |
| >3 | 39 | 436 | 369 | 4.12% | 6.14% | 3.65% |

According to table 3, it can be found that the word number is mainly 1, 2, and 3, and two words get the largest proportion, so how to combine candidate keyphrases is also a key question for keyphrase determination. In the process of keyphrase extraction by Topical PageRank algorithm, Liu et al. constructed noun phrases with pattern (adjective)*(noun)+ which denotes zero or more adjectives followed by one or more nouns as their candidate keyphrases, and then selected keyphrases by its weight value [43]. This article selects the noun phrases ending pattern used in Liu et al [43]. Then, we calculated the weight value of the candidate phrases by the formula below:

$$Weight(p) = \frac{\sum_{i=1}^{n} Weight(w_i)}{n} \quad (3)$$

where $Weight(w_i)$ represents the weight of the $i-th$ word in candidate keyword groups $p$, n refers to the number of words in the phrase $p$. That is, the weight of the phrase $p$ is the average weight of all words in $p$.

### 3.3.2 Automatic Keyphrase Extraction Methods Based on Supervised Learning

**(1) Naïve Baye**

Naïve Baye(NB) algorithm is based on the Bayes theorem and assumes that features are mutually independent, which is used for solving the classification problems. Given sample feature x, the probability that the sample belongs to category y is:

$$P(y|x) = \frac{P(x|y)P(y)}{P(x)} \quad (4)$$

In this article, considering TF*IDF value and the first appearance position, we adopt the extraction tool -- KEA[3] (Keyphrase Extraction Algorithm) to finish the task.

**(2) Conditional Random Field**

CRF is a discriminative probability model based on an undirected graph proposed by Lafferty et al. [17]. Given a random variable $X$, if each random variable $Y_v$ obeys Markov attribute, and then $(X,Y)$ constitutes a conditional random field. According to the basic theory of CRF, when G is a first-order linear list and simultaneously given a sequence of observations $X = \{x_1, x_2, \cdots, x_n\}$, the probability distribution $P(Y|X)$ of annotation sequence $Y = \{y_1, y_2, \cdots, y_n\}$ expressed as [17]:

$$P(y|x) = \frac{1}{Z(x)} \exp \sum_i \sum_k \lambda_k f_k(y_{i-1}, y_i, x, i) \quad (5)$$

where $Z(x)$ is the normalized factor, $f_k(y_{i-1}, y_i, x, i)$ is the eigenfunction between the marked position $i-1$ and $i$ of observed sequence, and $\lambda_k$ denotes the weight of the eigenfunction. In this article, we use CRF++[4] to realize the keyphrase sequence annotation and utilize 5-tag set[35] $\{key_S, key_B\ , key_M, key_E, key_N\}$ to annotate the training set. The specific definition of the tags is shown in Table 4. Given the relevant content $X$ of an academic paper, which can be represented as the word sequence $X$=$\{X_1, X_2\ , \ldots, X_{|X_i|}\ \}$, where $|X_i|$ is the length of content $X$. And the target is to produce a tag sequence $<Y_1, Y_2\ , \ldots, Y_{|X_i|}>$, where $Y_{|X_i|}$ represents whether $X_{|X_i|}$ is the part of keyphrase.

[3] http://community.nzdl.org/kea/index.html

Table 4: Definition of each tag $y_{i,|x_i|}$

| Tag | Definition |
|---|---|
| ***Key_S*** | $X_i$ represents a one-word keyphrase. |
| ***Key_B*** | $X_i$ denotes the first word of a keyphrase. |
| ***Key_M*** | $X_i$ is part of a keyphrase but it is neither the first nor the last word of the keyphrase. |
| ***Key_E*** | $X_i$ represents the last word of a keyphrase. |
| ***Key_N*** | $X_i$ is not a keyphrase or part of a keyphrase. |

The CRF model requires feature engineering support. Considering the factors that affect sequence annotation and the characteristics of our datasets, we selected nine text features, as shown in Table 5. Since each feature may not improve the performance of the model, so we need to select the best features for keyphrase extraction, which is introduced in Section 4.1.

Table 5: Text features for keyphrase extraction

| Name | Description |
|---|---|
| ***F1*** | *Part of speech* |
| ***F2*** | *Length of word* |
| ***F3*** | *Location of the first occurrence* |
| ***F4*** | *Word frequency in the full text* |
| ***F5*** | *Word frequency in reference title* |
| ***F6*** | *whether a phrase appears in title* |
| ***F7*** | *whether a phrase appears in reference title* |
| ***F8*** | *TF*IDF* |
| ***F9*** | *TextRank* |

**(3) BiLSTM-CRF**

BiSLTM-CRF is a deep learning model, as well as a sequence labeling model, which is often used in information extraction tasks, e.g. automatic keyphrase extraction (AKE) [18], named entity recognition (NER) [44]. Figure 1 is the model structure of the BiLSTM-CRF in our experiment. It mainly contains three layers, i.e., input layer, hidden layer, and output layer.

**Input layer:** We use word embedding to obtain the vector representation of words. In the experiment, we not only use word-level encoding vector, but also use character-level encoding vector that obtained by BiLSTM module. It can help the model identify new words. Besides, we also added word-level text features, as shown in Table 5. Given an input sentence s=$\{X_1,X_2,\ldots,X_n\}$, the input word vector of each word $X_i$ consists of word-level word vector $V_i$, character-level word vector $V_w$ and word-level text features $V_t$, which is sent to the hidden layer.

**Hidden layer:** It is a combination of the BiLSTM and CRF layers. After the outputs of BiSLTM, we again added text features to each word encoding, which obtains the semantic representation of each word and enhances the performance of the model.

**Output layer:** The output of the output layer is the classification label of each word. In the experiment, we used the BIO labeling method, i.e., ‘B’ represents the first word of the keyphrase or a one-word keyphrase. ‘I’ represents the words except for the first word of the keyphrase, ‘O’ is not a keyphrase or part of a keyphrase.

[4] https://taku910.github.io/crfpp/

In the experiment, we divided each paper into multiple sentences and then labeled each sentence according to the keyphrase assigned by the author. During training, we feed sentence-label pairs to the model using the training set. Then, we use the test set to evaluate its performance. Appendix 1 lists the hyper-parameters configuration of the model.

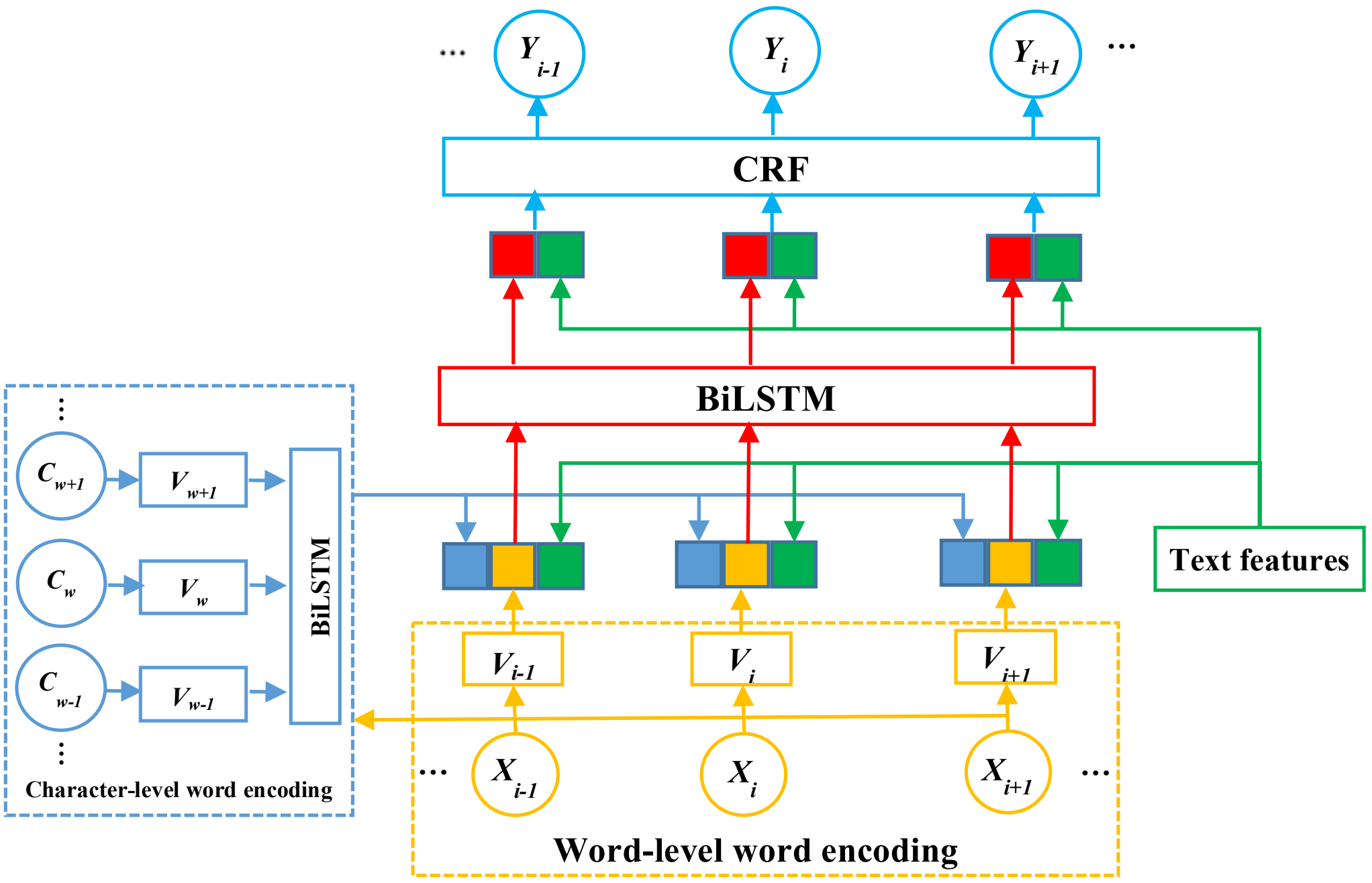

Figure 2: The model structure of the BiLSTM-CRF in the experiment

**Note:** $X_i$ represents the word that appears in the i-th position in a sentence. $C_w$ represents the character that appears in the w-th position in a word.

### 3.4 Evaluation Metrics

The keyphrases returned through different methods are divided into two situations: the keyphrases extracted are those assigned by the author or not. This article mainly adopts three evaluation metrics, namely, accuracy, recall and micro F-Measure ($F_1$). The calculation formulas are as follows:

$$P = \frac{TP}{TE} \tag{6}$$

$$R = \frac{TP}{TA} \tag{7}$$

$$F_1 = \frac{2 \times P \times R}{P + R} \tag{8}$$

Where $TP$ is the total number of correct keyphrases extracted by a method, $TE$ denotes the total number of automatic extracted keyphrases, $TA$ represents the total number of keyphrases assigned by the author.

## 4 Experiment and Result Analysis

In order to explore the influence of introducing reference information on keyphrase extraction effect,

this article selectes TF*IDF, TextRank as unsupervised methods and NB, CRF and BiLSTM-CRF as supervised methods, and used SemEval-2010, PubMed and LIS-2000 for the experiment. In this section, we first perform feature selection analysis to explore the influence of different text features on two supervised models, i.e. CRF and BiLSTM. Then, we use the keyphrase extraction models to evaluate the quality of multiple corpora constructed using different logical structure texts, especially focusing on the role of adding reference titles.

### 4.1 Feature selection analysis

The nine text features shown in Table 5 are selected for the CRF and BiLSTM models to extract keyphrases, but not every feature can play a role, so feature selection is required. In order to evaluate the performance of individual features, we first use all the features to extract keyphrases on the titles and abstracts of the datasets SemEval-2010, PubMed and LIS-2000, and then remove a feature in turn to see whether the $F_1$ value changes. If the $F_1$ value rises, the current feature has a negative effect, and vice versa. The results are shown in Table 6.

Table 6: The performance of individual features on the titles and abstracts of the datasets SemEval-2010, PubMed and LIS-2000

| Features \ Dataset | SemEval-2010 | | PubMed | | LIS-2000 | |
|---|---|---|---|---|---|---|
| | **CRF** | **BiLSTM-CRF** | **CRF** | **BiLSTM-CRF** | **CRF** | **BiLSTM-CRF** |
| **All** | **9.57** | **18.06** | **18.45** | **19.17** | **22.24** | **19.33** |
| **-$F_1$** | **8.93**$^+$ | **15.58**$^+$ | **17.27**$^+$ | 20.37 | **22.23**$^+$ | **18.97**$^+$ |
| **-$F_2$** | 11.11 | **14.84**$^+$ | **16.92**$^+$ | 19.60 | 22.72 | **18.80**$^+$ |
| **-$F_3$** | 12.02 | **15.73**$^+$ | **17.47**$^+$ | 19.29 | **22.07**$^+$ | **18.70**$^+$ |
| **-$F_4$** | 10.04 | **14.57**$^+$ | **17.58**$^+$ | 19.59 | 23.65 | 19.74 |
| **-$F_5$** | 10.67 | **16.24**$^+$ | **17.26**$^+$ | 19.71 | **22.16**$^+$ | **18.65**$^+$ |
| **-$F_6$** | **7.08**$^+$ | **17.09**$^+$ | **17.89**$^+$ | **18.91**$^+$ | **22.20**$^+$ | **18.35**$^+$ |
| **-$F_7$** | 9.65 | **17.54**$^+$ | **18.02**$^+$ | 19.76 | 22.47 | **19.30**$^+$ |
| **-$F_8$** | **7.21**$^+$ | **16.86**$^+$ | **17.48**$^+$ | **19.16**$^+$ | 23.28 | **18.22**$^+$ |
| **-$F_9$** | 11.91 | 18.38 | 18.59 | 19.87 | 22.58 | **18.65**$^+$ |
| **Best features** | **$F_{1,6,8}$** | **$F_{1,2,3,4,5,6,7,8}$** | **$F_{1,2,3,4,5,6,7,8}$** | **$F_{6,8}$** | **$F_{1,3,5,6}$** | **$F_{1,2,3,5,6,7,8,9}$** |

Note: The "+" sign in the table indicates that the feature has a positive effect.

It can be seen from Table 6 that the feature $F_6$ (whether a phrase appears in title) has a positive effect on extracting keyphrases for CRF and BiLSTM-CRF models on all datasets. The feature $F_1$ (part of speech) has no effect in extracting keyphrases for the BiLSTM-CRF model on the PubMed dataset. But it has a positive effect in other cases. The feature $F_8$ (TF*IDF) is similar to $F_1$. Besides, we found that the two features $F_5$ (word frequency in reference title) and $F_7$ (whether a phrase appears in reference title) related to reference information also have a positive effect when used in CRF or BiLSTM-CRF models on different datasets.

In subsequent experiments, we select the best feature set for CRF and BiLSTM-CRF to extract keyphrases. Although these best feature sets are selected on the corpus of title and abstract, they can still be used on other corpora. In addition, the two features $F_8$ (TF*IDF) and $F_9$ (TextRank) can not help CRF and BiLSTM-CRF models to extract more correct keyphrases in some cases, but they are often used for keyphrase extraction in unsupervised methods. Therefore, we take these two features as mandatory features and add them to the best feature set to extract keyphrases.

### 4.2 The influence of introducing reference titles on keyphrase extraction

Based on the title (T) and abstract (A) of the paper, we construct ten corpora for keyphrase extraction by adding different logical structure texts, i.e. introduction (I), conclusion (C), first sentence of paragraph (Fp), last sentence of paragraph (Lp), reference title (R) and full text (F). Then, we use

TF*IDF, TextRank, NB, CRF and BiLSTM-CRF to extract keyphrases from the constructed ten corpora. Finally, we use four evaluation metrics, i.e. precision, recall, $F_1$ values and running time (training and testing time) to conduct a detailed analysis.

It can be seen from Tab le 8-11 that the performance of supervised keyphrase extraction methods are better than that of unsupervised keyphrase extraction methods. In many cases, when the reference titles are introduced, the performance of the five keyphrase extraction methods is better than that without introducing reference titles, and it consumes less training or testing time. In the subsequent section, we respectively analyze the unsupervised learning methods and the supervised learning methods with or without introducing reference titles in detail. Besides, T-test is used to enhance the credibility of the experimental results.

### 4.2.1 Keyphrase Extraction Based on Unsupervised Learning with Reference Titles

In the experiment with TF*IDF and TextRank algorithms, we firstly perform the data preprocessing on ten corpora of SemEval-2010, PubMed and LIS-2000, and then conduct candidate keyphrase weight calculation, candidate keyphrase combination, keyphrase determination and results evaluation respectively. According to the number of keyphrases given by author, each academic paper is given 3.877, 5.398 and 5.052 unique keyphrases on average in the SemEval-2010, PubMed and LIS-2000, respectively. Thus, the precision, recall and $F_1$ values of top 3, 5, 7 and 10 keyphrases are calculated and shown in Table 8-10.

By comparing the results of Corpus_TA in Table 8-10, it can be seen that the addition of reference title has a beneficial impact on keyphrase extraction in PubMed and LIS-2000. It can extract more correct keyphrases than adding introduction and conclusion, first sentence of paragraph, last sentence of paragraph and the full text. In the dataset of SemEval-2010, when the number of extractions is 5, 7 and 10, the three evaluation metrics of Corpus_TAR are superior to those of other corpora in the TextRank algorithm. Besides, when the number of extractions is 3 and 10, the performance of TF*IDF algorithm with introducing reference titles is better than that without introducing reference titles. However, in some cases, adding reference titles is not better than the performance of Corpus_TA. This phenomenon appears more frequently in the dataset of SemEval-2010. The reason is that introducing reference titles can expand the corpus of keyphrase extraction and change the relative weight of the keyphrases in the corpus. It can improve the performance of keyphrase extraction, but it may also reduce the performance. Therefore, introducing reference titles to enhance the extraction quality with TF*IDF and TextRank are effective within a certain range, which is better than adding other logical structure texts.

### 4.2.2 Keyphrase Extraction Based on Supervised Learning with Reference titles

Based on supervised learning algorithms or models, i.e. NB, CRF, BiLSTM-CRF, we extract the keyphrases from ten corpora of SemEval-2010, PubMed and LIS-2000. The evaluation metrics including precision, recall and $F_1$ value of extraction results are presented in Table 8-10. In the CRF and BiLSTM-CRF models, the keyphrase extraction task is regarded as a sequence labeling task, which makes the extracted keyphrases have no weight values, so there is no need to select the top N keyphrases for evaluation based on the weights.

In NB algorithm, when extracting the same top N keyphrases, the Corpus_TAR containing the reference titles has higher precision, recall and $F_1$ values than the Corpus_TA containing only titles and abstracts, which indicates that the reference title has a positive impact on the keyphrase extraction. Although the current result may not be the best one, the time consumed for model training and testing is the least. In CRF model, the three evaluation metrics of Corpus_TAR are superior to those of Corpus_TA. In the dataset of PubMed, the Corpus_TAR containing the reference titles achieves the best performance on BiLSTM-CRF model. However, it performs poorly in SemEval-2010 and LIS-2000. The reason is that it is difficult to optimize its hyperparameters on all datasets. Therefore, in the experiment, we used the same hyperparametric configuration shown in Appendix 1 for all datasets, which makes the current configuration applicable to the dataset of PubMed, but not to the datasets of SemEval-2010 and LIS-2000. Moreover, we also find that adding the first sentence or the last sentence of the paragraph can also make the performance of the CRF and BiLSTM-CRF models reach the best in some cases. Compared with adding full text, adding reference titles helps to extract more correct keyphrases based on the titles and abstracts. In short, introducing reference title information has a positive effect on the keyphrase extraction performance in the supervised learning methods.

### 4.3 The statistical significance test with or without reference titles

According to the previous analysis results, adding reference titles to the title and abstract corpus (Corpus_TAR) can improve the performance of the keyphrase extraction methods. In order to enhance the credibility of the experimental results, we refer to the method of Awan et al. [45], and use T-test to calculate the statistical significance to understand the improvement of the results. In the process of establishing the hypothesis, the null hypothesis is that the $F_1$ value of each model trained with Corpus_TA is not different from the $F_1$ value of each model trained with Corpus_TAR, and the alternative hypothesis is that there is a difference between them. The $\alpha$ value is set to 0.05. If the calculated p-value is less than 0.05, we can reject the null hypothesis and vice versa. It can be seen in Table 7 that the p-values of all datasets are less than 0.05, which clearly shows that we can reject the null hypothesis of the results on all datasets. Therefore, we believe that the performance of the model trained by adding reference titles to the title and abstract (Corpus_TAR) is different than that of the model trained using only the title and abstract (Corpus_TA), and the performance of the former is better than that of the latter. The reliability of this conclusion can reach 95%.

Table 7: T-test results on Corpus_TA and Corpus_TAR using $F_1$ values in SemEval-2010, PubMed and LIS-2000

| Item \ Dataset | SemEval-2010 | | PubMed | | LIS-2000 | |
|---|---|---|---|---|---|---|
| | **TA** | **TAR** | **TA** | **TAR** | **TA** | **TAR** |
| 1 | 8.15 | 10.03 | 10.18 | 11.71 | 8.03 | 9.02 |
| 2 | 10.12 | 10.12 | 11.42 | 12.95 | 9.12 | 10.21 |
| 3 | 10.57 | 10.18 | 12.09 | 12.94 | 9.26 | 10.42 |
| 4 | 9.47 | 10.38 | 11.62 | 12.98 | 9.38 | 10.17 |
| 5 | 8.78 | 7.52 | 9.28 | 10.18 | 8.78 | 10.82 |
| 6 | 9.16 | 10.12 | 10.19 | 10.99 | 9.61 | 11.55 |
| 7 | 7.83 | 11.35 | 10.38 | 10.93 | 9.72 | 12.16 |
| 8 | 7.63 | 9.77 | 10.54 | 10.81 | 9.94 | 11.63 |
| 9 | 21.32 | 21.94 | 16.82 | 18.91 | 14.52 | 15.08 |
| 10 | 19.28 | 24.58 | 18.59 | 22.41 | 14.72 | 16.75 |
| 11 | 18.00 | 21.92 | 19.08 | 22.34 | 14.53 | 16.63 |
| 12 | 14.84 | 18.32 | 17.95 | 21.05 | 13.69 | 15.99 |
| 13 | 17.28 | 22.88 | 18.45 | 20.99 | 23.84 | 24.78 |
| 14 | 18.06 | 17.57 | 19.42 | 19.44 | 19.74 | 19.37 |
| **P-value** | **0.004*** | | **0.000*** | | **0.000*** | |

**Note**: ∗. Significant at α=0.05

Table 8: Keyphrase extraction performance of multiple corpora constructed using different logical structure texts on the dataset of SemEval-2010

| **Algorithms/ Models** | **TOP N** | **TA** | | | **TAI** | | | **TAC** | | | **TAFp** | | | **TALp** | | | **TAR** | | | **TAF** | | | **TAFR** | | | **TAICFpLp** | | | **TAICFpLpR** | | |
|---|---|---|---|---|---|---|---|---|---|---|---|---|---|---|---|---|---|---|---|---|---|---|---|---|---|---|---|---|---|---|---|
| | | **P** | **R** | **$F_1$** | **P** | **R** | **$F_1$** | **P** | **R** | **$F_1$** | **P** | **R** | **$F_1$** | **P** | **R** | **$F_1$** | **P** | **R** | **$F_1$** | **P** | **R** | **$F_1$** | **P** | **R** | **$F_1$** | **P** | **R** | **$F_1$** | **P** | **R** | **$F_1$** |
| **TF*IDF** | **3** | 9.03 | 7.43 | 8.15 | 9.03 | 7.43 | 8.15 | **11.81** | **9.71** | **10.66** | 9.03 | 7.43 | 8.15 | 6.94 | 5.71 | 6.27 | 11.11 | 9.14 | 10.03 | 5.56 | 4.57 | 5.02 | 5.56 | 4.57 | 5.02 | 6.94 | 5.71 | 6.27 | 6.25 | 5.14 | 5.64 |
| | **5** | 8.75 | 12.00 | 10.12 | 7.92 | 10.86 | 9.16 | **10.42** | **14.29** | **12.05** | 9.17 | 12.57 | 10.60 | 5.42 | 7.43 | 6.27 | 8.75 | 12.00 | 10.12 | 6.25 | 8.57 | 7.23 | 5.83 | 8.00 | 6.75 | 5.42 | 7.43 | 6.27 | 5.42 | 7.43 | 6.27 |
| | **7** | 8.04 | 15.43 | 10.57 | 7.44 | 14.29 | 9.78 | **8.04** | **15.43** | **10.57** | 7.44 | 14.29 | 9.78 | 4.76 | 9.14 | 6.26 | 7.74 | 14.86 | 10.18 | 5.06 | 9.71 | 6.65 | 5.06 | 9.71 | 6.65 | 5.65 | 10.86 | 7.44 | 5.95 | 11.43 | 7.83 |
| | **10** | 6.46 | 17.71 | 9.47 | 6.67 | 18.29 | 9.77 | 6.67 | 18.29 | 9.77 | 6.25 | 17.14 | 9.16 | 4.38 | 12.00 | 6.41 | **7.08** | **19.43** | **10.38** | 4.38 | 12.00 | 6.41 | 4.58 | 12.57 | 6.72 | 4.58 | 12.57 | 6.72 | 5.00 | 13.71 | 7.33 |
| **TextRank** | **3** | 9.72 | 8.00 | 8.78 | 6.25 | 5.14 | 5.64 | **10.42** | **8.57** | **9.40** | 9.03 | 7.43 | 8.15 | 5.56 | 4.57 | 5.02 | 8.33 | 6.86 | 7.52 | 5.56 | 4.57 | 5.02 | 5.56 | 4.57 | 5.02 | 6.94 | 5.71 | 6.27 | 6.25 | 5.14 | 5.64 |
| | **5** | 7.92 | 10.86 | 9.16 | 5.42 | 7.43 | 6.27 | 8.33 | 11.43 | 9.64 | 7.92 | 10.86 | 9.16 | 4.17 | 5.71 | 4.82 | **8.75** | **12.00** | **10.12** | 4.17 | 5.71 | 4.82 | 3.75 | 5.14 | 4.34 | 4.58 | 6.29 | 5.30 | 5.00 | 6.86 | 5.78 |
| | **7** | 5.95 | 11.43 | 7.83 | 5.36 | 10.29 | 7.05 | 6.85 | 13.14 | 9.00 | 6.85 | 13.14 | 9.00 | 4.46 | 8.57 | 5.87 | **8.63** | **16.57** | **11.35** | 4.17 | 8.00 | 5.48 | 4.46 | 8.57 | 5.87 | 4.76 | 9.14 | 6.26 | 4.76 | 9.14 | 6.26 |
| | **10** | 5.21 | 14.29 | 7.63 | 5.42 | 14.86 | 7.94 | 5.83 | 16.00 | 8.55 | 5.83 | 16.00 | 8.55 | 3.54 | 9.71 | 5.19 | **6.67** | **18.29** | **9.77** | 4.17 | 11.43 | 6.11 | 4.79 | 13.14 | 7.02 | 4.58 | 12.57 | 6.72 | 5.21 | 14.29 | 7.63 |
| **NB** | **3** | 23.61 | 19.43 | 21.32 | 24.31 | 20.00 | 21.94 | 22.22 | 18.29 | 20.06 | 24.31 | 20.00 | 21.94 | 25.00 | 20.57 | 22.57 | 24.31 | 20.00 | 21.94 | 24.31 | 20.00 | 21.94 | 25.00 | 20.57 | 22.57 | **27.08** | **22.29** | **24.45** | 25.69 | 21.14 | 23.20 |
| | **5** | 16.67 | 22.86 | 19.28 | 20.42 | 28.00 | 23.61 | 17.08 | 23.43 | 19.76 | **21.25** | **29.14** | **24.58** | 18.33 | 25.14 | 21.20 | **21.25** | **29.14** | **24.58** | 19.17 | 26.29 | 22.17 | 19.17 | 26.29 | 22.17 | 20.42 | 28.00 | 23.61 | **21.25** | **29.14** | **24.58** |
| | **7** | 13.69 | 26.29 | 18.00 | 16.37 | 31.43 | 21.53 | 14.29 | 27.43 | 18.79 | **17.56** | **33.71** | **23.09** | 14.88 | 28.57 | 19.57 | 16.67 | 32.00 | 21.92 | 16.37 | 31.43 | 21.53 | 16.67 | 32.00 | 21.92 | 16.37 | 31.43 | 21.53 | 16.96 | 32.57 | 22.31 |
| | **10** | 10.17 | 27.43 | 14.84 | 12.71 | 34.86 | 18.63 | 11.88 | 32.57 | 17.40 | 13.33 | 36.57 | 19.54 | 11.25 | 30.86 | 16.49 | 12.50 | 34.29 | 18.32 | 12.50 | 34.29 | 18.32 | 12.92 | 35.43 | 18.93 | 12.50 | 34.29 | 18.32 | **13.96** | **38.29** | **20.46** |
| **CRF** | **-** | 30.88 | 12.00 | 17.28 | 24.79 | 16.57 | 19.86 | 27.47 | 14.29 | 18.80 | 22.95 | 16.00 | 18.86 | 22.86 | 13.71 | 17.14 | **32.29** | 17.71 | **22.88** | 20.00 | 19.43 | 17.71 | 16.74 | **21.14** | 18.69 | 20.75 | 18.86 | 19.76 | 17.71 | 19.43 | 18.53 |
| **BiLSTM-CRF** | **-** | 20.74 | **16.00** | 18.06 | 16.95 | 11.43 | 13.65 | 17.36 | 12.00 | 14.19 | **23.28** | 15.43 | **18.56** | 20.78 | 9.14 | 12.70 | 21.49 | 14.86 | 17.57 | 17.54 | 11.43 | 13.84 | 22.68 | 12.57 | 16.18 | 20.43 | 10.86 | 14.18 | 18.75 | 10.29 | 13.28 |

**Note**: The yellow, green and blue bold fonts in the table represent the largest of the P, R and $F_1$ value obtained from different corpora using the same model, respectively.

Table 9: Keyphrase extraction performance of multiple corpora constructed using different logical structure texts on the dataset of PubMed

| Algorithms/ Models | TOP N | TA | | | TAI | | | TAC | | | TAFp | | | TALp | | | TAR | | | TAF | | | TAFR | | | TAICFpLp | | | TAICFpLpR | | |
|---|---|---|---|---|---|---|---|---|---|---|---|---|---|---|---|---|---|---|---|---|---|---|---|---|---|---|---|---|---|---|---|
| | | P | R | $F_1$ | P | R | $F_1$ | P | R | $F_1$ | P | R | $F_1$ | P | R | $F_1$ | P | R | $F_1$ | P | R | $F_1$ | P | R | $F_1$ | P | R | $F_1$ | P | R | $F_1$ |
| TF*IDF | 3 | 14.27 | 7.91 | 10.18 | 15.66 | 8.68 | 11.17 | 13.26 | 7.35 | 9.46 | 14.77 | 8.19 | 10.54 | 12.88 | 7.14 | 9.19 | **16.41** | **9.10** | **11.71** | 10.61 | 5.88 | 7.56 | 10.98 | 6.09 | 7.83 | 13.01 | 7.21 | 9.28 | 12.75 | 7.07 | 9.10 |
| | 5 | 11.89 | 10.99 | 11.42 | 12.58 | 11.62 | 12.08 | 10.76 | 9.94 | 10.33 | 12.20 | 11.27 | 11.71 | 11.52 | 10.64 | 11.06 | **13.48** | **12.46** | **12.95** | 8.56 | 7.91 | 8.22 | 9.39 | 8.68 | 9.02 | 10.23 | 9.45 | 9.82 | 10.45 | 9.66 | 10.04 |
| | 7 | 10.72 | 13.86 | 12.09 | 10.94 | 14.14 | 12.33 | 9.58 | 12.39 | 10.81 | 10.34 | 13.37 | 11.66 | 9.79 | 12.67 | 11.05 | **11.47** | **14.84** | **12.94** | 7.47 | 9.66 | 8.42 | 8.28 | 10.71 | 9.34 | 8.98 | 11.62 | 10.13 | 9.31 | 12.04 | 10.50 |
| | 10 | 8.96 | 16.52 | 11.62 | 9.71 | 17.91 | 12.60 | 8.23 | 15.19 | 10.68 | 8.90 | 16.45 | 11.55 | 8.26 | 15.26 | 10.72 | **10.00** | **18.47** | **12.98** | 6.55 | 12.11 | 8.50 | 7.27 | 13.44 | 9.44 | 7.99 | 14.77 | 10.37 | 8.26 | 15.26 | 10.72 |
| TextRank | 3 | 13.01 | 7.21 | 9.28 | 13.26 | 7.35 | 9.46 | 10.86 | 6.02 | 7.74 | 11.99 | 6.65 | 8.55 | 10.23 | 5.67 | 7.29 | **14.27** | **7.91** | **10.18** | 8.59 | 4.76 | 6.12 | 9.22 | 5.11 | 6.57 | 9.85 | 5.46 | 7.02 | 10.73 | 5.95 | 7.65 |
| | 5 | 10.61 | 9.80 | 10.19 | 11.36 | 10.50 | 10.91 | 8.94 | 8.26 | 8.58 | 10.23 | 9.45 | 9.82 | 8.71 | 8.05 | 8.37 | **11.44** | **10.57** | **10.99** | 7.05 | 6.51 | 6.77 | 7.35 | 6.79 | 7.06 | 9.32 | 8.61 | 8.95 | 9.55 | 8.82 | 9.17 |
| | 7 | 9.20 | 11.90 | 10.38 | **9.69** | **12.53** | **10.93** | 8.34 | 10.78 | 9.40 | 8.82 | 11.41 | 9.95 | 7.90 | 10.22 | 8.91 | **9.69** | **12.53** | **10.93** | 6.22 | 8.05 | 7.02 | 6.66 | 8.61 | 7.51 | 7.95 | 10.29 | 8.97 | 7.85 | 10.15 | 8.85 |
| | 10 | 8.13 | 14.98 | 10.54 | 8.12 | 14.98 | 10.53 | 6.94 | 12.81 | 9.00 | 7.54 | 13.93 | 9.78 | 7.08 | 13.09 | 9.19 | **8.33** | **15.40** | **10.81** | 5.72 | 10.57 | 7.42 | 5.80 | 10.71 | 7.52 | 6.74 | 12.46 | 8.75 | 6.70 | 12.39 | 8.70 |
| NB | 3 | 23.75 | 13.02 | 16.82 | 25.86 | 14.28 | 18.39 | 24.33 | 13.44 | 17.31 | 24.75 | 13.72 | 17.65 | 23.61 | 13.09 | 16.84 | 26.52 | 14.70 | 18.91 | 26.14 | 14.49 | 18.64 | **26.77** | **14.84** | **19.09** | 24.75 | 13.72 | 17.65 | 25.76 | 14.28 | 18.37 |
| | 5 | 19.48 | 17.77 | 18.59 | 21.06 | 19.38 | 20.19 | 21.75 | 20.01 | 20.85 | 21.06 | 19.45 | 20.23 | 20.98 | 19.38 | 20.15 | **23.33** | **21.55** | **22.41** | 21.74 | 20.08 | 20.88 | 22.58 | 20.85 | 21.68 | 22.12 | 20.43 | 21.24 | 23.11 | 21.34 | 22.19 |
| | 7 | 17.03 | 21.69 | 19.08 | 18.14 | 23.37 | 20.43 | 17.76 | 22.88 | 20.00 | 18.29 | 23.65 | 20.63 | 18.40 | 23.79 | 20.75 | 19.81 | 25.61 | 22.34 | 19.32 | 24.98 | 21.79 | 20.24 | 26.17 | 22.83 | 18.78 | 24.28 | 21.18 | **20.29** | **26.24** | **22.89** |
| | 10 | 13.96 | 25.12 | 17.95 | 15.38 | 28.27 | 19.93 | 14.54 | 26.73 | 18.83 | 15.15 | 27.99 | 19.66 | 15.04 | 27.78 | 19.52 | 16.22 | 29.95 | 21.05 | 16.29 | 30.09 | 21.14 | 16.78 | 31.00 | 21.77 | 15.76 | 29.11 | 20.45 | **17.05** | **31.49** | **22.12** |
| CRF | - | **24.59** | 14.77 | 18.45 | 23.57 | 19.31 | 21.23 | 23.60 | 18.82 | 20.94 | 24.06 | **21.48** | **22.70** | 21.67 | 18.54 | 19.98 | 21.90 | 20.15 | 20.99 | 18.91 | 17.88 | 18.38 | 17.74 | 18.87 | 18.29 | 19.01 | 17.56 | 18.25 | 18.92 | 19.28 | 19.09 |
| BiLSTM-CRF | - | 22.51 | 17.07 | 19.42 | 27.71 | 12.53 | 17.25 | **30.75** | 11.41 | 16.64 | 20.04 | 15.68 | 17.59 | 22.59 | 14.28 | 17.50 | 20.09 | **18.82** | **19.44** | 12.24 | 14.77 | 13.38 | 15.32 | 16.59 | 15.93 | 21.75 | 15.12 | 17.84 | 15.99 | 18.40 | 17.11 |

**Note**: The yellow, green and blue bold fonts in the table represent the largest of the P, R and $F_1$ value obtained from different corpora using the same model, respectively.

Table 10: Keyphrase extraction performance of multiple corpora constructed using different logical structure texts on the dataset of LIS-2000

| Algorithms/ Models | TOP N | TA | | | TAI | | | TAC | | | TAFp | | | TALp | | | TAR | | | TAF | | | TAFR | | | TAICFpLp | | | TAICFpLpR | | |
|---|---|---|---|---|---|---|---|---|---|---|---|---|---|---|---|---|---|---|---|---|---|---|---|---|---|---|---|---|---|---|---|
| | | P | R | $F_1$ | P | R | $F_1$ | P | R | $F_1$ | P | R | $F_1$ | P | R | $F_1$ | P | R | $F_1$ | P | R | $F_1$ | P | R | $F_1$ | P | R | $F_1$ | P | R | $F_1$ |
| TF*IDF | 3 | 10.83 | 6.39 | 8.03 | 11.17 | 6.58 | 8.28 | 10.42 | 6.14 | 7.73 | 7.83 | 4.62 | 5.81 | 9.58 | 5.65 | 7.11 | **12.17** | **7.17** | **9.02** | 7.67 | 4.52 | 5.69 | 7.33 | 4.32 | 5.44 | 9.08 | 5.35 | 6.74 | 9.17 | 5.40 | 6.80 |
| | 5 | 9.20 | 9.04 | 9.12 | 9.35 | 9.18 | 9.27 | 8.95 | 8.79 | 8.87 | 7.40 | 7.27 | 7.33 | 7.50 | 7.37 | 7.43 | **10.30** | **10.12** | **10.21** | 6.65 | 6.53 | 6.59 | 6.40 | 6.29 | 6.34 | 7.55 | 7.42 | 7.48 | 7.35 | 7.22 | 7.28 |
| | 7 | 8.00 | 11.00 | 9.26 | 8.14 | 11.20 | 9.43 | 7.79 | 10.71 | 9.02 | 6.79 | 9.33 | 7.86 | 6.79 | 9.33 | 7.86 | **9.00** | **12.38** | **10.42** | 5.71 | 7.86 | 6.62 | 5.82 | 8.01 | 6.74 | 6.61 | 9.09 | 7.65 | 6.61 | 9.09 | 7.65 |
| | 10 | 7.07 | 13.90 | 9.38 | 6.83 | 13.41 | 9.05 | 6.42 | 12.62 | 8.52 | 6.22 | 12.23 | 8.25 | 5.73 | 11.25 | 7.59 | **7.67** | **15.08** | **10.17** | 4.95 | 9.72 | 6.56 | 5.12 | 10.07 | 6.79 | 5.80 | 11.39 | 7.69 | 5.78 | 11.35 | 7.65 |
| TextRank | 3 | 11.83 | 6.97 | 8.78 | 11.25 | 6.63 | 8.34 | 10.50 | 6.19 | 7.79 | 9.08 | 5.35 | 6.74 | 9.42 | 5.55 | 6.98 | **14.58** | **8.60** | **10.82** | 8.33 | 4.91 | 6.18 | 9.42 | 5.55 | 6.98 | 9.00 | 5.30 | 6.67 | 11.25 | 6.63 | 8.34 |
| | 5 | 9.70 | 9.53 | 9.61 | 9.55 | 9.38 | 9.46 | 8.95 | 8.79 | 8.87 | 7.85 | 7.71 | 7.78 | 8.20 | 8.06 | 8.13 | **11.65** | **11.44** | **11.55** | 7.95 | 7.81 | 7.88 | 7.90 | 7.76 | 7.83 | 8.05 | 7.91 | 7.98 | 9.00 | 8.84 | 8.92 |
| | 7 | 8.39 | 11.54 | 9.72 | 8.11 | 11.15 | 9.39 | 7.64 | 10.51 | 8.85 | 6.71 | 9.23 | 7.78 | 7.25 | 9.97 | 8.40 | **10.50** | **14.44** | **12.16** | 6.89 | 9.48 | 7.98 | 6.79 | 9.33 | 7.86 | 7.04 | 9.68 | 8.15 | 7.57 | 10.41 | 8.77 |
| | 10 | 7.50 | 14.73 | 9.94 | 6.58 | 12.92 | 8.71 | 6.53 | 12.82 | 8.65 | 6.25 | 12.28 | 8.28 | 6.35 | 12.48 | 8.42 | **8.77** | **17.24** | **11.63** | 5.75 | 11.30 | 7.62 | 5.92 | 11.64 | 7.85 | 6.33 | 12.43 | 8.38 | 6.55 | 12.87 | 8.68 |
| NB | 3 | 19.58 | 11.54 | 14.52 | 18.83 | 11.10 | 13.97 | 18.75 | 11.05 | 13.91 | 19.17 | 11.30 | 14.22 | 18.33 | 10.81 | 13.60 | 20.33 | 11.98 | 15.08 | 19.50 | 11.49 | 14.46 | 19.75 | 11.64 | 14.65 | 19.33 | 11.39 | 14.34 | **20.67** | **12.18** | **15.33** |
| | 5 | 14.86 | 14.59 | 14.72 | 15.55 | 15.28 | 15.41 | 14.80 | 14.54 | 14.67 | 15.80 | 15.52 | 15.66 | 15.20 | 14.93 | 15.06 | **16.90** | **16.60** | **16.75** | 15.70 | 15.42 | 15.56 | 16.15 | 15.86 | 16.01 | 16.05 | 15.77 | 15.91 | 16.40 | 16.11 | 16.25 |
| | 7 | 12.55 | 17.24 | 14.53 | 13.25 | 18.22 | 15.35 | 12.25 | 16.85 | 14.19 | 13.75 | 18.91 | 15.92 | 13.25 | 18.22 | 15.34 | **14.37** | **19.74** | **16.63** | 13.14 | 18.07 | 15.22 | 13.64 | 18.76 | 15.80 | 13.46 | 18.52 | 15.59 | 13.82 | 19.01 | 16.00 |
| | 10 | 10.35 | 20.19 | 13.69 | 11.17 | 21.91 | 14.80 | 10.44 | 20.48 | 13.83 | 11.20 | 22.00 | 14.84 | 10.85 | 21.32 | 14.38 | **12.07** | **23.67** | **15.99** | 11.10 | 21.81 | 14.71 | 11.53 | 22.64 | 15.28 | 11.28 | 22.15 | 14.94 | 11.62 | 22.84 | 15.41 |
| CRF | - | **29.76** | 19.89 | 23.84 | 27.60 | 23.87 | 25.60 | 27.64 | 21.74 | 24.32 | 27.30 | 24.21 | 25.66 | 28.48 | 23.92 | **26.00** | 26.87 | 22.99 | 24.78 | 19.45 | 23.69 | 21.35 | 19.05 | 24.53 | 21.44 | 20.78 | 23.22 | 21.91 | 19.97 | **24.66** | 22.06 |
| BiLSTM-CRF | - | 20.35 | **19.16** | **19.74** | 21.69 | 15.23 | 17.89 | 20.89 | 16.75 | 18.59 | 18.11 | 17.83 | 17.97 | 18.89 | 16.21 | 17.45 | 20.67 | 18.22 | 19.37 | **22.43** | 13.16 | 16.59 | 21.06 | 16.80 | 18.69 | 20.84 | 15.18 | 17.56 | 22.26 | 14.69 | 17.70 |

**Note**: The yellow, green and blue bold fonts in the table represent the largest of the P, R and $F_1$ value obtained from different corpora using the same model, respectively.

Table 11: The multiple of training or testing time of ten corpora based on training or testing time of the Corpus_TA

| Dataset | Algorithms/ Models | TA | | TAI | | TAC | | TAFp | | TALp | | TAR | | TAF | | TAFR | | TAICFpLp | | TAICFpLpR | |
|---|---|---|---|---|---|---|---|---|---|---|---|---|---|---|---|---|---|---|---|---|---|
| | | Train | Test | Train | Test | Train | Test | Train | Test | Train | Test | Train | Test | Train | Test | Train | Test | Train | Test | Train | Test |
| SemEval-2010 | TF*IDF | - | 1.0 | - | 8.0 | - | 3.9 | - | 17.5 | - | 20.4 | - | 2.5 | - | 161.8 | - | 172.5 | - | 74.2 | - | 82.5 |
| | TextRank | - | 1.0 | - | 4.2 | - | 2.6 | - | 6.5 | - | 7.4 | - | 2.0 | - | 29.0 | - | 28.9 | - | 18.2 | - | 19.4 |
| | NB | 1.0 | 1.0 | 1.9 | 1.6 | 1.4 | 1.4 | 2.7 | 2.3 | 2.8 | 2.2 | 1.2 | 1.2 | 8.0 | 5.8 | 8.1 | 4.7 | 6.0 | 3.8 | 7.1 | 3.7 |
| | CRF | 1.0 | 1.0 | 8.2 | 4.2 | 5.0 | 2.6 | 16.7 | 6.2 | 17.3 | 7.2 | 2.5 | 1.8 | 81.9 | 25.2 | 89.5 | 25.6 | 58.0 | 17.0 | 59.4 | 19.2 |
| | BiLSTM-CRF | 1.0 | 1.0 | 5.6 | 1.7 | 2.4 | 1.2 | 9.0 | 2.2 | 9.7 | 2.3 | 3.2 | 1.3 | 38.1 | 6.5 | 41.2 | 6.6 | 23.8 | 4.3 | 25.9 | 4.6 |
| PubMed | TF*IDF | - | 1.0 | - | 4.5 | - | 5.6 | - | 7.8 | - | 7.9 | - | 9.5 | - | 67.7 | - | 88.0 | - | 30.5 | - | 50.7 |
| | TextRank | - | 1.0 | - | 2.3 | - | 2.8 | - | 3.3 | - | 3.2 | - | 2.9 | - | 17.6 | - | 15.5 | - | 9.3 | - | 10.6 |
| | NB | 1.0 | 1.0 | 2.6 | 1.8 | 2.2 | 1.5 | 2.7 | 1.7 | 2.6 | 1.6 | 2.5 | 1.6 | 9.1 | 4.4 | 13.4 | 7.0 | 7.0 | 3.5 | 9.0 | 4.5 |
| | CRF | 1.0 | 1.0 | 4.4 | 2.7 | 3.5 | 3.3 | 5.6 | 4.1 | 5.9 | 3.8 | 4.3 | 3.3 | 16.8 | 66.7 | 20.7 | 76.5 | 12.6 | 45.8 | 14.1 | 55.4 |
| | BiLSTM-CRF | 1.0 | 1.0 | 3.0 | 2.0 | 3.3 | 2.3 | 4.1 | 2.6 | 3.6 | 2.4 | 4.7 | 2.9 | 17.5 | 9.5 | 21.3 | 11.4 | 11.1 | 6.2 | 14.7 | 8.1 |
| LIS-2000 | TF*IDF | - | 1.0 | - | 6.7 | - | 4.0 | - | 10.6 | - | 10.8 | - | 6.8 | - | 105.1 | - | 128.2 | - | 47.0 | - | 56.7 |
| | TextRank | - | 1.0 | - | 2.9 | - | 2.3 | - | 4.2 | - | 4.4 | - | 3.1 | - | 17.7 | - | 22.1 | - | 11.1 | - | 13.3 |
| | NB | 1.0 | 1.0 | 3.0 | 1.2 | 1.7 | 1.0 | 2.9 | 1.4 | 3.0 | 1.3 | 1.8 | 1.0 | 11.8 | 4.0 | 12.1 | 4.3 | 7.7 | 2.7 | 9.0 | 3.2 |
| | CRF | 1.0 | 1.0 | 4.5 | 3.9 | 4.3 | 2.7 | 7.3 | 5.4 | 8.9 | 5.5 | 3.4 | 2.6 | 24.1 | 87.1 | 26.7 | 94.4 | 15.5 | 55.6 | 18.9 | 62.2 |
| | BiLSTM-CRF | 1.0 | 1.0 | 3.2 | 2.5 | 2.7 | 2.1 | 5.3 | 3.8 | 5.2 | 3.8 | 4.0 | 2.9 | 21.1 | 14.6 | 24.1 | 16.4 | 13.7 | 9.5 | 16.6 | 11.0 |

**Note**: For example, in the datatset of SemEval-2010, the test time of TF*IDF algorithm on Corpus_TAR is 2.5 times that on Corpus_TA.

## 5 Discussion

Utilizing the reference information for keyphrase extraction have been used earlier in the literature. For example, Krapivin et al. proposed the SVM+NLP method in their research, and extracted keyphrases by using the title, references and sections headers of the paper, which achieved good performance, but lacked the comparison of keyphrase extraction results from other sections of the paper [46]. Bhaskar et al. used the reference word as a feature of the CRF model to extract keyphrases from scientific articles, but they did not evaluate the effectiveness of this feature in detail [47]. In response to these shortcomings, this paper analyzes the influence of reference information on the keyphrase extraction task from two aspects: the text features related to reference information and the use of reference titles for keyphrase extraction.

**The text features related to reference information:** We evaluated the impact of the nine text features shown in Table 5 on the CRF and BiLSTM-CRF models. We found that two text features related to reference information, i.e. word frequency in reference title and whether a phrase appears in reference title, can help CRF and BiLSTM-CRF models extract more keyphrases. It indicates that the reference word features are useful for keyphrase extraction, which coincides with krapivin et al.

**The use of reference titles for keyphrase extraction:** In order to explore the influence of using reference titles for keyphrase extraction, we take the titles and abstracts of academic papers as the basic corpus, and construct ten corpora by adding different logical structure texts to extract keyphrases. It is found that the performance of the model trained with the corpus containing titles, abstracts and reference titles is better. Prior to this, Nguyen et al. also conducted a similar research, but they ignored the reference titles for keyphrase extraction [48]. Therefore, we consider the influence of adding reference titles on the performance of the model based on the title, abstract, introduction, conclusion, first sentence of paragraph and last sentence of paragraph of the papers. We found that adding reference titles has the potential to improve model performance. Although we did not use the best training corpus they mentioned for keyphrase extraction, the results of our report still have practical significance.

## 6 Conclusion and Future Work

For the sake of exploring the impact of introducing reference title information on the keyphrase extraction task, this article adopts five methods to compare the extraction results with or without reference title information, which shows that the quality of the keyphrase extraction can be improved to some extent by adding reference titles. In this experiment, it can be seen that the effect of keyphrase extraction with supervised learning methods is better than that with unsupervised learning methods, but the two types of keyphrase extraction methods indeed support the effectiveness of reference title information putting in keyphrase extraction task.

We have explored the influence of introducing reference information on automatic keyphrase extraction in academic papers through five methods of automatic keyphrase extraction and the experimental results prove the value of reference title information, but there are also some shortcomings. In this article, all the reference titles are directly incorporated into the corpus for keyphrase extraction, and there may be some references that are not related to the paper's topic. Thus, we can try to calculate the text similarity between the paper and its corresponding references first, and then choose the reference information with high similarity. This article verifies the effectiveness of reference title information for keyphrase extraction, moreover, we can question that whether citing literature will affect the quality of keyphrase extraction and try to add citing literature information to expand the context in the future.

## Acknowledgement

This work is supported by National Natural Science Foundation of China (Grant No.72074113), Open Fund Project of Fujian Provincial Key Laboratory of Information Processing and Intelligent Control (Minjiang University) (No. MJUKF-IPIC201903) and Practice Innovation Program of Jiangsu Province (No. KYCX21_0426).

## Appendix 1: Hyper-parameters configuration of BiLSTM-CRF model

| Parameters | Value |
|---|---|
| Word embedding dimension | 128 |
| Character embedding dimension | 32 |
| Maximum word length | 32 |
| Maximum sentence length | 256 |
| Hidden layer dimensions | 128 |
| Number of layers of LSTM | 1 |
| Droupout rate | 0.5 |
| Epochs | 20 |
| Batch size | 128 |
| Learning rate | 1e-2 |
| Optimizer algorithm | Adam |
| Weight decay for optimizer | 1e-4 |
| Scheduler factor for optimizer | 0.1 |
| Scheduler patience for optimizer | 1 |